\documentclass[10pt]{iopart}
\usepackage{epsf}
\usepackage{graphicx}
\usepackage{epstopdf}

\usepackage{amssymb}
\usepackage{graphicx}

\newcommand{\norm}[1]{\|\, #1 \,\|}
\newcommand{\h}{\mathcal{H}}
\newcommand{\A}{{\bf A}}
\newcommand{\HA}{H_{\bf A}}

\newtheorem{thm}{Theorem}[section]
\newtheorem{cor}[thm]{Corollary}
\newtheorem{lem}[thm]{Lemma}

\begin{document}

\title{Norm estimates of complex symmetric operators applied to quantum systems}

\author{Emil Prodan}
\address{PRISM, Princeton University, Princeton, NJ 08544}

\author{Stephan R. Garcia and Mihai Putinar}
\address{Mathematics Department,
 University of California,
  Santa
Barbara, CA 93106-3080}

\date{today}

\begin{abstract}
This paper communicates recent results in theory of complex
symmetric operators and shows, through two non-trivial examples,
their potential usefulness in the study of Schr\"odinger operators.
In particular, we propose a formula for computing the norm of a compact 
complex symmetric operator. This observation is applied to two concrete
problems related to quantum mechanical systems. First, we give sharp
estimates on the exponential decay of the resolvent and the
single-particle density matrix for Schr\"odinger operators with
spectral gaps. Second, we provide new ways of evaluating the
resolvent norm for Schr\"odinger operators appearing in the complex
scaling theory of resonances.
\end{abstract}


\maketitle

\section{Introduction}

About half a century ago, Glazman  laid the foundations for the
theory of unbounded complex symmetric operators \cite{Glazman1,
Glazman}. Since then, his fundamental ideas have been successfully
tested on several classes of differential operators \cite{CG,
Knowles, Race}. Recently, two of the authors discovered an
additional structure in the polar decomposition of a complex
symmetric operator \cite{GarciaPutinar2}. For certain unbounded
operators with compact resolvent, the refined polar decomposition
leads to a new method for estimating the norm of their resolvent. In
the present note, we exploit this idea in conjunction with the
complex scaling method for Schr\"odinger operators.

Although Quantum Mechanics is built on the theory of selfadjoint
operators, one must often deal with non-selfadjoint operators. For
instance, this is the case when appealing to the complex scaling
technique. This method became a standard tool in the theory of
Schr\"odinger operators and turned out to be the key to several
problems such as: the absence of singular continuous spectrum
\cite{AguilarCombes, BalslevCombes}, the calculus of resonances and
the convergence of time-dependent perturbation theory \cite{Simon73},
and asymptotic behavior of the eigenvectors \cite{CombesThomas}. As
the examples in this paper show, the complex scaling technique
naturally leads to complex symmetric operators.

Our first application deals with Schr\"odinger operators with a
spectral gap. We provide sharp exponential decay estimates on the
resolvent and the single-particle density matrix. Such estimates
became increasingly important since it was realized that the
localization of the single-particle density matrix provides the key
to efficient numerical electronic structure algorithms for systems
with large number of particles \cite{Wu02}. For 1D periodic
insulators, exact exponential decays can be derived from Kohn's
analytic results \cite{Kohn59}. In an attempt to generalize these
results to higher dimensions, des Cloizeaux \cite{Cloizeaux64a,
Cloizeaux64b} developed a method which can be regarded as the first
application of the complex scaling idea. He proved the exponential
decay of the single-particle density matrix for a class of 3D
insulators. Relatively recently, we have seen a renewed interest in
the subject and remarkable new exact results in dimensions higher
than one \cite{BeAr, HeVa, JeKro,  TaFry}. These results, however,
are limited to periodic systems and some of them to the extreme
tight-binding limit. In the present note, we treat the general case
of gapped Schr\"odinger operators, which find applications, in
addition to the periodic insulators, to amorphous insulators,
molecular liquids, or large molecules.

For periodic systems, it has long been known that the exponential decay
of the resolvent, when the energy is in the gap and close to the gap
edges, is proportional to square root of the distance from $E$ to
the gap edge. The theory of the effective mass \cite{Kohn57}
provides a simple way of estimating this exponential decay constant.
Relatively recently, it was proven that this qualitative behavior is
present in any gapped systems \cite{Barbaroux, Hislop}. The present
paper sharpens these previous estimates. The goal is to find the
best quantitative estimate of the exponential decay constant, with
the energy spectrum as the only input.

In our second application, we show that the technique of estimating
norms of complex symmetric operators extends to certain operators
with non-compact resolvent, such as the complex scaled Hamiltonians
from the problem of resonances.

\section{Complex symmetric operators}
This section is a brief account of relevant results (both old and new)  about
complex symmetric operators. For full details and examples the
reader can consult \cite{GarciaPutinar1, GarciaPutinar2}.

We first consider bounded operators. Let $\h$ denote a separable
complex Hilbert space which carries a \emph{conjugation}
$C:\h\longrightarrow\h$, an \emph{antilinear} operator
satisfying the conditions $C^2 = I$ and $\langle Cf,Cg \rangle =  \langle g,f \rangle$ for all $f,g \in \h$.
A bounded operator $T$ on $\h$ is called \emph{$C$-symmetric} if $T = CT^*C$. More
generally, $T$ is called \emph{complex symmetric} if there exists a
$C$ such that $T$ is $C$-symmetric.  The terminology arises from the fact
that $T$ is complex symmetric if and only if it has a symmetric matrix representation
with respect to some orthonormal basis \cite{GarciaPutinar1}.

Examples of bounded complex symmetric operators include normal
operators, Hankel operators, finite Toeplitz matrices, Jordan model
operators (the infinite dimensional analogs of Jordan blocks), the
Volterra integration operator, and several other classes as well
(see \cite{GarciaPutinar1,GarciaPutinar2}). Some of these classes of
operators appeared in direct applications like atomic collisions
\cite{Brown} and bound states problem in periodic chains
\cite{Prunele}.

The following simple factorization theorem (from \cite{GarciaPutinar2})
is the main ingredient in the proofs contained throughout this note.

\begin{thm}\cite{GarciaPutinar2} If
$T:\h\longrightarrow\h$ is a bounded $C$-symmetric operator, then
there exists a conjugation $J:\h\longrightarrow\h$, which
commutes with the spectral measure of $|T| = \sqrt{T^*T}$, such
that $T = CJ|T|$.
\end{thm}

\noindent The theorem above asserts the equivalence of the
\emph{antilinear} eigenvalue problems
\begin{equation}\label{equivalence}
Tf = \lambda Cf \ \ \Leftrightarrow \ \ |T|f = \lambda Jf.
\end{equation}
We may assume that $\lambda$ is real and positive, since we
may multiply either of these equations by a suitable unimodular
constant. This equivalence is important for the following reason.
The norm of a compact operator $T$ is equal to the largest
eigenvalue of $|T|$. However, working with $|T|$ instead of the
original operator $T$ introduces unwanted and potentially
serious complications. Eq.~(\ref{equivalence}) provides the
following convenient alternative: the norm of a compact
$C$-symmetric operator $T$ is equal to the largest positive solution $\lambda$ of the
\emph{antilinear} eigenvalue problem, $Tf = \lambda Cf$.  This is 
because the auxiliary conjugation $J$ fixes an orthonormal basis of each 
spectral subspace of $|T|$.

We now turn to the case of unbounded operators. A densely defined,
closed operator $T$ is $C$-\emph{symmetric} if $T \subset CT^*C$ and
$C$-\emph{selfadjoint} if $T = CT^*C$, i.e. their domains coincide
and, on this common domain, they are equal.

The study of unbounded complex symmetric operators was pioneered by
Glazman \cite{Glazman1, Glazman}, who established a complex
symmetric parallel to von Neumann's theory of selfadjoint extensions
of symmetric operators, although certain classes of $C$-symmetric
operators had appeared earlier in von Neumann's work \cite{vN}. A
renewed interest in Glazman's theory was sparked by its application
to certain Dirac-type operators \cite{CG} and the realization that
the closely related class of \emph{$C$-unitary} operators is
relevant to the study of complex scaling transformations in quantum
mechanics \cite{Riss}. Moreover, certain Sturm-Liouville operators
with complex potentials can also be treated similarly \cite{Knowles,
Race}. Further examples are furnished by Schr\"odinger operators
$-\Delta + v$ with \emph{complex potentials} $v$ (where $C$ is
simply complex conjugation) subject to appropriate boundary
conditions \cite{Glazman, Race}.  One can also consider
Schr\"odinger operators $-\Delta + v$ with real potentials $v$, but
complex (non-selfadjoint) two point boundary conditions, in which
case the conjugation $C$ is slightly more involved.

From the classical theory of selfadjoint operators, one knows that a
symmetric operator has selfadjoint extensions if and only if its
deficiency indices are the same. In contrast, \emph{every}
$C$-symmetric unbounded operator admits a $C$-selfadjoint extension
$\widetilde{T}$. Unfortunately, not all unbounded $C$-selfadjoint
operators possess a spectral resolution and a corresponding fine
functional calculus. Nevertheless, if an unbounded $C$-selfadjoint
operator has a compact resolvent, then a canonically associated
antilinear eigenvalue problem always has a complete set of mutually
orthogonal eigenfunctions:

\begin{thm}\cite{GarciaPutinar2}
 If $T: \mathcal{D}(T) \longrightarrow H$ is an
 unbounded $C$-selfadjoint operator with compact resolvent
 $(T-z)^{-1}$ for some complex number $z$, then there exists an
 orthonormal basis $(u_n)_{n=1}^\infty$ of $\h$ consisting of
 solutions of the antilinear eigenvalue problem:
 $$
   (T-z)u_n = \lambda_n Cu_n
$$
 where $(\lambda_n)_{n=1}^{\infty}$ is an increasing sequence
 of positive numbers tending to $\infty$.
\end{thm}
\noindent This result is a direct consequence of the refined polar
decomposition $T = CJ|T|$ for \emph{bounded} $C$-symmetric operators
described in Theorem 2.1. Our main technical tool in estimating the
norms of resolvents of certain unbounded operators is contained in
the following corollary:

\begin{cor}\cite{GarciaPutinar2}
  If $T$ is a densely-defined $C$-selfadjoint operator
  with compact resolvent
  $(T-z)^{-1}$ for some complex number $z$, then
  \begin{equation}\label{ResolventEstimate}
    \norm{(T - z)^{-1}} = \frac{1}{\inf_n \lambda_n}
  \end{equation}
  where the $\lambda_n$ are the positive solutions to the
  antilinear eigenvalue problem:
 \begin{equation}\label{Anti}
   (T-z)u_n = \lambda_n Cu_n.
 \end{equation}
 Here $\| \ \|$ denotes the operator norm: $\|A\|\equiv
\sup_{\|\phi\|=1}\sqrt{\langle A\phi,A\phi\rangle}$.
\end{cor}

We also remark that the refined polar decomposition $T= CJ|T|$
applies, under certain circumstances, to unbounded $C$-selfadjoint
operators.  The following extension will play a central role in our second
application:

\begin{thm}\cite{GarciaPutinar2}
  If $T: \mathcal{D}(T) \longrightarrow \h$ is a densely
  defined $C$-selfadjoint operator with zero in its resolvent,
  then $T = CJ|T|$ where $|T|$ is a positive selfadjoint operator
  (in the von Neumann sense) satisfying
  $\mathcal{D}(|T|) = \mathcal{D}(T)$ and $J$ is a conjugation
  operator on $\h$ which commutes with the spectral measure
  of $|T|$. Conversely, any operator of the form described above is
  $C$-selfadjoint.
\end{thm}

We close this section with a few remarks about the computation 
of the eigenvalues $\lambda_n = \lambda_n(|T|)$ and associated
eigenfunctions via the antilinear problem (\ref{equivalence}) or (\ref{Anti}). 
If $T$ is a complex $C$-symmetric operator, then the bilinear (as opposed to sesquilinear) form
$$ [Tx, y] = \langle Tx, Cy \rangle = \langle |T|x, Jy \rangle$$
can detect these values via a min-max principle
analogous to the corresponding procedure for selfadjoint operators 
(see \cite[XIII.1]{ReedSimon}).
More specifically, if $\lambda_0 \geq \lambda_1 \geq \lambda_2 \geq \cdots \geq 0$ 
are the eigenvalues of the compact positive operator $|T|$, arranged 
in decreasing order and repeated according to multiplicity, then the 
evenly indexed eigenvalues $\lambda_{2n}$ are given by the 
following variational principle:
\begin{equation}\label{Minimax}
  \lambda_{2n}= \min_{{\rm codim} V = n}\, \max_{u \in V, \| u \| = 1} \Re [Tu,u].
\end{equation}
The same principle applied to a rank one perturbation $T + \xi \otimes \xi$ (where $\xi$
is any vector satisfying $\|\xi\| > \|T\|$) yields the odd eigenvalues.
The proof of (\ref{Minimax}), along with several applications,
can be found in \cite{DGP}.  We also remark that (\ref{Minimax})
generalizes the recently discovered min-max principle for symmetric 
matrices with complex entries \cite{Danciger}.

If one is only concerned with $\norm{T}$, then simply take $n = 0$ in (\ref{Minimax}) 
and note that
$$\norm{T} = \max_{u \in \h,\, \| u \| = 1} \Re [Tu,u].$$
The antilinear eigenproblem $Tf = \lambda Cf$ and min-max principle (\ref{Minimax})
can be broken into real linear problems and also have the advantage that they
do not involve the direct calculation of $T^*T$.  Moreover, one 
might potentially find numerical approximations to the $\lambda_n$ via
procedures similar to those in \cite[XIII.1]{ReedSimon}.

\section{Exponential Decay of the Resolvent for Gapped Systems}
In this section, we consider the problem of finding sharp estimates
on the exponential decay of the resolvent for Schr\"odinger
operators with a gap in the spectrum. A short account on the subject
has been already given in the Introduction.

We now formulate the problem and the main result. Let $-{\bf
\nabla}_D^2$ denote the Laplace operator with zero (Dirichlet)
boundary conditions over a finite domain (with smooth boundary)
$\Omega \subset \mathbf{R}^d$; let $v({\bf x})$ be a scalar
potential, which is ${\bf \nabla}_D^2$-relatively bounded, with
relative bound less than one. Throughout this section, all potentials
$v$ are presumed to be bounded from below. By measuring the energy
from the bottom of the potential, we can assume without loss of
generality that $v({\bf x}) \geq 0$. We also include a magnetic field, 
described by a smooth vector potential $\A({\bf x})$. The total Hamiltonian is
$$\HA:{\mathcal D}({\bf \nabla}_D^2) \longrightarrow L^2(\Omega),\ \
\HA=-({\bf \nabla}+i\A)^2+v({\bf x}).$$ 
The assumption on $H_\A$ is that
its energy spectrum $\sigma$ consists of two parts,
$\sigma\subset[0,E_-]\cup[E_+,\infty)$, which are separated by a gap
$G\equiv E_+-E_->0$. We refer to the spectrum $\sigma_\pm$
above/below the gap as as the upper/lower band. The corresponding
spectral projectors are denoted by $P_\pm$.

Let $E\in(E_-,E_+)$ and $G_E=(\HA-E)^{-1}$ be the resolvent. We are
interested in the behavior of the kernel $G_E({\bf x},{\bf y})$ for
large separations $|{\bf x}-{\bf y}|$. Instead of looking directly
at the pointwise behavior, we take the average
$$  \bar{G}_E({\bf x}_1,{\bf x}_2)\equiv\frac{1}{\omega_\epsilon^2}
    \int\limits_{|{\bf x}-{\bf x}_1|\leq \epsilon}d{\bf x}
    \int\limits_{|{\bf y}-{\bf x}_2|\leq \epsilon}d{\bf y} \
    G_E({\bf x},{\bf y}),$$
where $\omega_\epsilon$ is the volume of a sphere of radius
$\epsilon$ in $\mathbf{R}^d$. The main result of this section is
stated below.

\begin{thm}
For $q$ smaller than a critical value $q_c(E)$, there exists
a constant $C_{q,E}$, independent of $\Omega$, such that:
\begin{equation}\label{UpperBound}
    |\bar{G}_E({\bf x}_1,{\bf x}_2)|\leq C_{q,E}e^{-q|{\bf x}_1-{\bf
    x}_2|}.
\end{equation}
$C_{q,E}$ is given by:
\begin{equation}
    C_{q,E}=\frac{\omega_\epsilon^{-1} e^{2q\epsilon}}{\min|E_\pm-E-q^2|}\cdot  \frac{1}{1-q/F(q,E)}
\end{equation}
with
\begin{equation}
    F(q,E)=\sqrt{\frac{(E_+-E-q^2)(E-E_-+q^2)}{4E_-}}.
\end{equation}
The critical value $q_c(E)$ is the positive solution of the equation
$q=F(q,E)$.
\end{thm}

\noindent{\bf Proof.} If $\chi_{{\bf x}}$ denotes the characteristic
function of the $\epsilon$ ball centered at ${\bf x}$ (i.e.
$\chi_{{\bf x}}({\bf x}^\prime)=1$ for $|{\bf x}^\prime-{\bf x}|\leq
\epsilon$ and $0$ otherwise), then one can equivalently write
$$\bar{G}_E({\bf x}_1,{\bf x}_2)=\omega_\epsilon^{-2}\langle
\chi_{{\bf x}_1},(\HA-E)^{-1}\chi_{{\bf x}_2}\rangle.$$ Given a vector
${\bf q}\in \mathbf{R}^d$ ($q\equiv|{\bf q}|$) of arbitrary
orientation and magnitude, let $U_{\bf q}$ denote the following
bounded and invertible map
$$U_{\bf q}:L^2(\Omega)\rightarrow L^2(\Omega), \ \ [U_{\bf
q}f]({\bf x})=e^{{\bf q}{\bf x}}f({\bf x}),$$ which leaves the
domain of $\HA$ unchanged. Let $H_{{\bf q},\A}\equiv U_{\bf q}\HA U_{\bf
q}^{-1}$ be the family of scaled Hamiltonians. Explicitly, they
are given by
\begin{equation}
    H_{{\bf q},\A}:{\mathcal D}({\bf \nabla}_D^2)\rightarrow L^2(\Omega),
    \ \ H_{{\bf q},\A}=\HA+2{\bf q}({\bf \nabla}+i\A)-q^2.
\end{equation}
We notice that for ${\bf q}\neq 0$, these are non-selfadjoint and are
not even complex symmetric operators (with respect to any natural
conjugation).  We also note that the identity
$$U_{\bf q}(\HA-E)^{-1}U_{\bf q}^{-1}=(H_{{\bf q},\A}-E)^{-1}$$
holds for all ${\bf q}\in \mathbf{R}^d$ (this happens only for
finite $\Omega$). If $\Omega=\mathbf{R}^d$, the identity holds as long as the
continuum spectrum (which moves as ${\bf q}$ is increased) does not step over $E$. We denote
\begin{equation}
    \gamma(q,E)=\sup\limits_{|{\bf q}|=q}\|(H_{{\bf q},\A}-E)^{-1}\|.
\end{equation}
As the following lines show, the entire problem can be reduced to
estimating $\gamma(q,E)$. Indeed, if $\varphi_1({\bf x}) \equiv
e^{-{\bf q}({\bf x}-{\bf x}_1)}\chi_{{\bf x}_1}({\bf x})$ and
$\varphi_2({\bf x}) \equiv e^{{\bf q}({\bf x}-{\bf x}_2)}\chi_{{\bf
x}_2}({\bf x})$, then
\begin{eqnarray*}
    |\bar{G}_E({\bf x}_1,{\bf x}_2)| &=&\omega_\epsilon^{-2}|\langle \varphi_1,(H_{{\bf q},\A}-E)^{-1}\varphi_2\rangle| e^{-{\bf q}({\bf x}_1-{\bf x}_2)}
    \\
    &\leq& \omega_\epsilon^{-1} e^{2q\epsilon}
    \gamma(q,E)e^{-{\bf q}({\bf x}_1-{\bf x}_2)}.
\end{eqnarray*}
Choosing ${\bf q}$ parallel to ${\bf x}_1-{\bf x}_2$, we infer at
this step that
$$|\bar{G}_E({\bf x}_1,{\bf x}_2)|\leq \omega_\epsilon^{-1} e^{2q\epsilon}
\gamma(q,E)e^{-q|{\bf x}_1-{\bf x}_2|}.$$

Next we estimate $\gamma(q,E)$ using the results on $C$-symmetric
operators presented in the previous section. As we already
mentioned, $H_{{\bf q},\A}$ is not complex symmetric. However, note
that the operators $H_{{\bf q},\A}$ and $H_{-{\bf q},\A}$ are dual:
$H^\ast_{{\bf q},\A}=H_{-{\bf q},\A}$.  Furthermore, if ${\mathcal C}$ denotes 
complex conjugation, then ${\mathcal C}H_{{\bf q},\A}=H_{{\bf
q},-\A}{\mathcal C}$. We define the following block-matrix operator
$\mathbf{H}$ and conjugation $C$ on $L^2(\Omega)\oplus L^2(\Omega)$:
$$
    \mathbf{H}=\left(%
    \begin{array}{cc}
      H_{{\bf q},\A} & 0 \\
      0 & H_{-{\bf q},-\A} \\
    \end{array}%
    \right),\ \
    C   = \left(%
    \begin{array}{cc}
      0 & {\mathcal C} \\
      {\mathcal C} & 0 \\
    \end{array}%
    \right).
$$
It is a simple task to check that $\mathbf{H}$ is $C$-selfadjoint:
$\mathbf{H}^\ast=C\mathbf{H}C$. Moreover,
\begin{equation}\label{NormEq}
    \|(\mathbf{H}-E)^{-1}\|=\|(H_{{\bf q},\A}-E)^{-1}\|=\|(H_{-{\bf
    q},-\A}-E)^{-1}\|.
\end{equation}
According to the previous section, the antilinear eigenvalue problem
(with $\lambda_n\geq0$)
\begin{equation}\label{EigP}
    (\mathbf{H}-E)\phi_n=\lambda_n C\phi_n
\end{equation}
generates an orthonormal basis $\phi_n$ in $L^2(\Omega)\oplus L^2(\Omega)$
and
\begin{equation}\label{bound}
    \|(\mathbf{H}-E)^{-1}\|=\frac{1}{\min_n \lambda_n}.
\end{equation}
If we write $\phi_n=f_n\oplus g_n$, the antilinear eigenvalue
problem Eq.~(\ref{EigP}), after we apply a complex conjugation on the second equation, is equivalent to
\begin{equation}\label{EigPP}
    \left\{%
    \begin{array}{c}
      (H_{{\bf q},\A}-E)f_n=\lambda_n \bar{g}_n \\
      (H_{-{\bf q},\A}-E)\bar{g}_n=\lambda_n f_n. \\
    \end{array}%
    \right.
\end{equation}
With $q$ small, such that $E+q^2$ lies in the spectral gap, the
polar decomposition
$$\HA-E-q^2=S|\HA-E-q^2|,$$ holds, where $S=P_+-P_-$. We take the scalar
product of the first equation in Eq.~(\ref{EigPP}) against the
vector $Sf_n$. Keeping only the real part of the result and solving
for $\lambda_n$, we find
\begin{equation}\label{LambdaEq}
    \lambda_n=\frac{|\langle f_n,|\HA-E-q^2|f_n\rangle
    +2\Re\langle Sf_n,{\bf q}({\bf \nabla}+i\A)f_n\rangle|}{|\Re\langle
    Sf_n,\bar{g}_n \rangle|}.
\end{equation}
After elementary manipulations, the second term in the numerator
can be rewritten as
$$
\Re\langle Sf_n,{\bf q}({\bf \nabla}+i\A)f_n\rangle
=2\Re\langle f_n,P_+[{\bf q}({\bf \nabla}+i\A)]P_-f_n\rangle.
$$
Moreover, denoting
\begin{equation}\label{Bq}
B_{\bf q}\equiv P_+|\HA-E-q^2|^{-1/2}[{\bf q}({\bf
\nabla}+i\A)]|\HA-E-q^2|^{-1/2}P_-,
\end{equation}
one obtains
$$
|\langle f_n,P_+[{\bf q}({\bf \nabla}+i\A)]P_-f_n\rangle| \leq
\frac{1}{2}\|B_{\bf q}\| \langle f_n,|\HA-E-q^2|f_n\rangle.
$$
Eq.~(\ref{LambdaEq}) implies
$$
\lambda_n\geq \min\{\,|E_\pm-E-q^2|\,\}(1-2\|B_{\bf
q}\|)\frac{\|f_n\|^2}{\|f_n\|\|g_n\|}.
$$
Similarly, by taking the scalar product of the second equation of
Eq.~(\ref{EigPP}) against $S\bar{g}_n$, one obtains:
$$
\lambda_n\geq \min\{\, |E_\pm-E-q^2| \,\}(1-2\|B_{\bf
q}\|)\frac{\|g_n\|^2}{\|f_n\|\|g_n\|}.
$$
The sum of the last two equations yields
\begin{equation}\label{boundl}
    \lambda_n\geq \min\{\, |E_\pm-E-q^2|\, \}(1-2\|B_{\bf q}\|).
\end{equation}

It remains to evaluate $\|B_{\bf q}\|$. Since the potential is
positive and we work with zero boundary conditions, the following
inequality between quadratic forms
$$
q^2(-({\bf \nabla}+i\A)^2+v+a)\geq [{\bf q}({\bf \nabla}+i\A)]^2,\ \ \forall \
a\geq0,
$$
holds true. Consequently
$$
\|[{\bf q}({\bf \nabla}+i\A)]|\HA+a|^{-1/2}\|\leq q,\ \ \forall \ a > 0.
$$
We can then insert $|\HA+a|^{-1/2}|\HA+a|^{1/2}$ after ${\bf q}({\bf
\nabla}+i\A)$ in Eq.~(\ref{Bq}) and use the above estimate and the
spectral theorem to get:
\begin{equation}\label{boundb}
    \|B_{\bf q}\|\leq q\sqrt{\frac{E_-+a}{(E_+-E-q^2)(E-E_-+q^2)}}.
\end{equation}
Finally one can pass to the limit $a\rightarrow 0$.
Eq.~(\ref{bound}), together with Eqs.~(\ref{boundl}) and
(\ref{boundb}) provide the desired estimate of
$\gamma(q,E)$.$\square$

\medskip

We have the following remarks. The reason we considered only finite domains is primarily
because the abstract machinery presented in the previous section applies only to
complex symmetric operators with compact resolvent. An extension to operators with non-compact resolvent is given in the next Section. However, since the constants in 
Theorem 3.1 are independent of the domain $\Omega$, the estimates remain valid
when $\Omega\rightarrow \mathbf{R}^d$, whenever this limit is well defined. 
For $E$ close to the gap
edges, a brute perturbation theory on $(H_{{\bf q},\A}-E)^{-1}$ leads
to an exponential decay constant proportional to $|E_\pm-E|$. The
correct behavior of $q_c(E)$ near the band edges is
$|E_\pm-E|^{1/2}$, as it was first shown in Ref.~\cite{Barbaroux}
and later reviewed in Ref.~\cite{Hislop}. Note that our estimate of
$q_c(E)$ does have the correct behavior near the gap edges.

\begin{cor} Consider the lower band completely filled.
Then the single-particle density matrix (i.e. the projector onto
the occupied states $P_-$) decays exponentially, with a rate
$\bar{q}$ satisfying
\begin{equation}\label{ExpDec}
    \bar{q}\geq\frac{G}{4\sqrt{E_-}}.
\end{equation}
\end{cor}

\noindent{\bf Proof.} Again, we look at the average
$$\bar{P}_-({\bf x}_1,{\bf x}_2)=\omega_\epsilon^{-2}\langle \chi_{{\bf
x}_1},P_-\chi_{{\bf x}_2}\rangle,$$ which has the following
representation:
$$
\bar{P}_-({\bf x}_1,{\bf x}_2)=\frac{i}{2\pi}\int_\Gamma
\bar{G}_E({\bf x}_1,{\bf x}_2)\,dE,
$$
where $\Gamma$ is a contour in the complex energy plane,
surrounding the lower band. The estimates given in the preceding
Theorem trivially extend to the case of complex energies:
$$
|\bar{G}_{E+i\zeta}({\bf x}_1,{\bf x}_2)|\leq C_{q,E}e^{-q|{\bf
x}_1-{\bf x}_2|},\ \ \forall \ q<q_c(E).
$$
Given that $\Gamma$ can be deformed so as to intersect the real axis
at any point in $(E_-,E_+)$, we need to find the energy where
$q_c(E)$ is maximum. We have
\begin{equation}
    q_c=\sqrt{\frac{(E_+-E-q_c^2)(E-E_-+q_c^2)}{4E_-}}\leq\frac{G}{4\sqrt{E_-}}
\end{equation}
with equality at energy
\begin{equation}
    \bar{E}=\frac{E_++E_-}{2}-\frac{G^2}{16E_-}.
\end{equation}
The lower bound of Eq.~(\ref{ExpDec}) is valid as long as $\bar{E}$
is in the gap.$\square$
\medskip

\begin{figure}
\begin{center}
  \includegraphics[width=8.6cm]{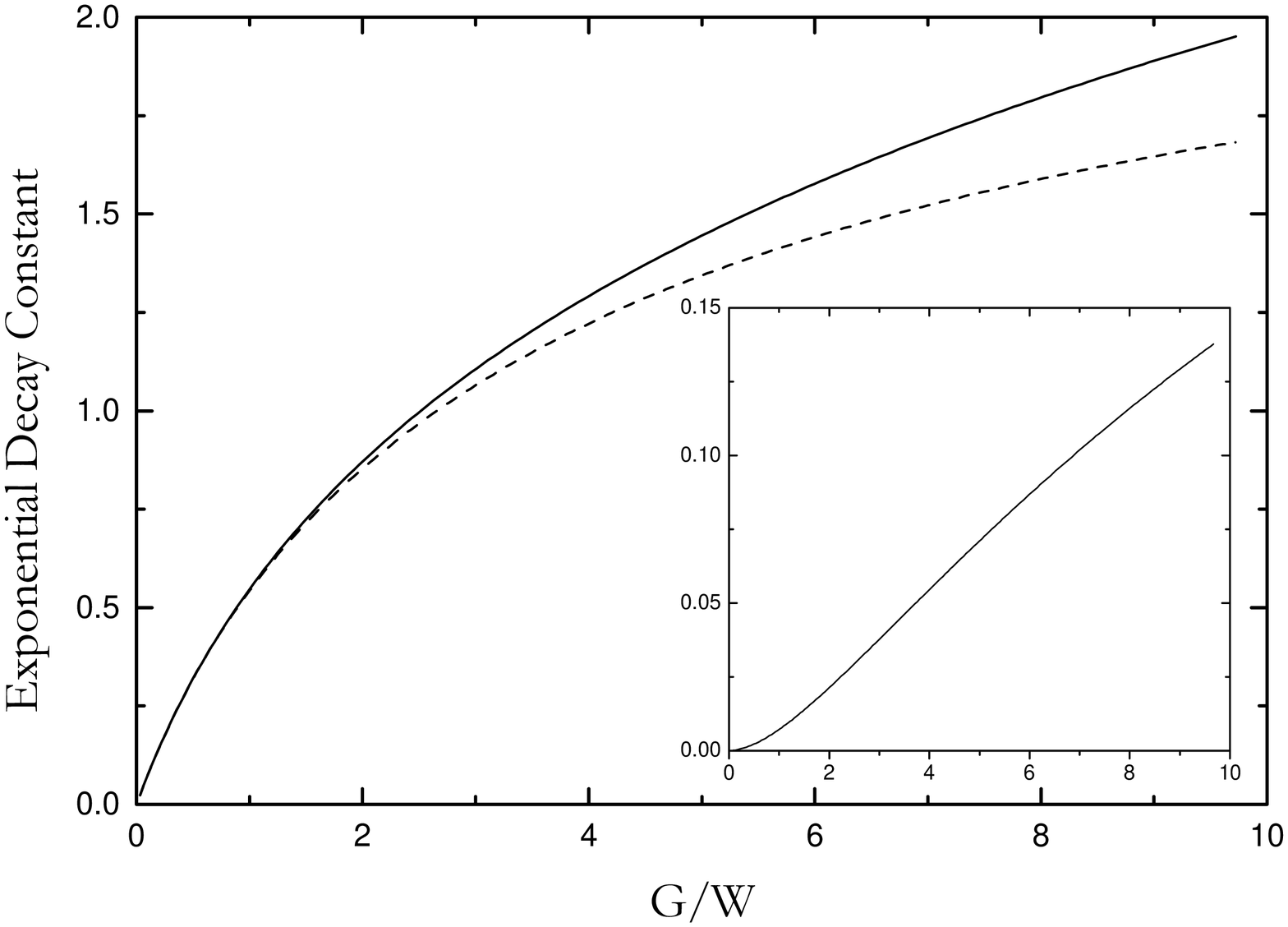}\\
  \end{center}
  \caption{The exponential decay constant of the single-particle
  density matrix as a function of $G/W$ for the Kronig-Penney
  insulator. The continuous line represents an exact calculation and
  the dashed line represents the estimate given in Eq.~(\ref{ExpDec}).
  The inset shows the relative difference between the two.
  }
\end{figure}

The lower bound on $\bar{q}$ given in Eq.~(\ref{ExpDec}) can be
calculated entirely from the energy spectrum. What is the best lower
bound on the exponential decay constant that one can get by only
using the information contained in the energy spectrum? We
illustrate by an example that Eq.~(\ref{ExpDec}) comes close to such
an optimal estimate. In Fig.~1, we consider a comparison between our
estimate Eq.~(\ref{ExpDec}) and the exact value of $\bar{q}$, for a
one dimensional insulator described by the Kronig-Penney model
\cite{Kronig}:
$$H=-\partial_x^2+v_0 \sum_n \delta(x-n),\ v_0>0,$$
with the first band completely filled. The Dirac-delta potential is not ${\bf \nabla}^2$
relatively bounded so the above example is not formally covered by Theorem 3.1. However, 
this potential is a relatively bounded form perturbation and our arguments can be extended to this larger class of potentials. By varying the strength of
the potential $v_0$, we sweep from a weak to strong insulating
regime, which we quantified by the ratio between the gap $G$ and the
width of the valence band $W$. One can see that even in the extreme
insulating regime (typically $G/W<5$), Eq.~(\ref{ExpDec}) estimates
the exponential decay to within a 15\% error. Notice that for
$G/W<5$, the error is less than 5\%.

\section{Norm Estimates on Resolvents Near Resonances}
In this section, we extend the results concerning norm estimates to
not necessarily compact resolvents of unbounded complex symmetric operators.
At the same time, we apply this technique to the problem of
locating the resonances of a specific class of Hamiltonians.

We first formulate the problem in precise terms. Let
$$
H:{\mathcal D}({\bf \nabla}^2)\longrightarrow L^2(\mathbf{R}^d),\ \
H=-{\bf \nabla}^2+v({\bf x})
$$
be a Hamiltonian with $v({\bf x})$ a dilation analytic potential in
a finite strip $|\Im\theta|<I_0$ and ${\bf
\nabla}^2$-relatively compact. We consider the usual dilation
operation:
$$[U(\theta)\psi]({\bf x})=e^{d \theta/2}\psi(e^\theta{\bf x})$$
and define the analytic family (of type A) of operators:
$$H_\theta \equiv U(\theta)HU(\theta)^{-1}=-e^{-2\theta}{\bf \nabla}^2+v(e^\theta {\bf x}),$$
where $\theta$ runs in the finite strip $|\Im\theta|<I_0$.
As a function of $\theta$, it is well known that
\cite{AguilarCombes, BalslevCombes, Simon72}:
\begin{enumerate}
  \item[(a)] the discrete spectrum $\sigma_d$ remains invariant,
  \item[(b)] the essential spectrum $\sigma_{ess}$ rotates down by an angle
    $-2\Im\theta$,
  \item[(c)] as the continuum rotates, it uncovers additional discrete spectrum
    (the resonances).
\end{enumerate}

In many practical situations, it is desired not only to locate the
resonances but also to know how they move under different
perturbations \cite{Nordlander,McCurdyBaer,McCurdyRescigno,Simon79}.
Here we are concerned with the second problem, where norm estimates
on the resolvent $(z-H_\theta)^{-1}$ for $z$ near the resonances
become especially important either for probing the stability of the
spectrum or for building perturbation series.

The Hamiltonians $H_\theta$ are $C$-selfadjoint relative to the
complex conjugation $Cf=\bar{f}$. The question that we want to
answer is if one can provide an exact norm estimate of
$(z-H_\theta)^{-1}$ for $z$ near a resonance, using the theory of
complex symmetric operators. The answer is contained in the
following theorem:

\begin{thm}
Let $\gamma w({\bf x})$ represent the change in $v({\bf x})$ and
$H(\gamma)=H+\gamma w$ denote the perturbed Hamiltonian. We assume
that both $v_\theta({\bf x})\equiv v(e^\theta {\bf x})$ and
$w_\theta({\bf x})\equiv w(e^\theta {\bf x})$ are ${\bf
\nabla}^2$-relatively bounded for $|\Im\theta|<I_0$, with
bound less than one. For $z$ close to a resonance $z_0$ of $H$ and
$\gamma$ sufficiently small, the following are true:
\begin{itemize}
    \item[{\it (i)}] $\sigma_{ess}(\,|H_\theta(\gamma)-z|\,)=[d(z,\theta),\infty)$.
    \item[{\it (ii)}] $\sigma_d(\,|H_\theta(\gamma)-z|\,)\cap
[0,d(z,\theta)] \neq \emptyset$.
    \item[{\it (iii)}] $\lambda_n\in \sigma_d(|H_\theta(\gamma)-z|)$ if and only
if there exists $\psi_n\in {\mathcal D}({\bf \nabla}^2)$ such that:
$$(H_\theta(\gamma)-z)\psi_n=\lambda_n C\psi_n.$$
Moreover $$\|(H_\theta(\gamma)-z)^{-1}\|=\frac{1}{\min_n \lambda_n}.$$
\end{itemize}

\noindent Above, $d(z,\theta)=|z\sin(2 \Im \theta-\alpha)|$
denotes the distance from $z$ to $\sigma_{ess}(H_\theta)$ (where
$z=|z|e^{-i\alpha}$).
\end{thm}

We require the following lemma:

\begin{lem}\label{LemmaMero}
Let $A$ and $B$ be two closed operators such that
$\mathcal{D}(A)\subset \mathcal{D}(B)$ and $B|A|^{-1}$ is compact.
Let $A+B$ be the closed sum on $\mathcal{D}(A)$. Then
$\sigma_{ess}|A+B|=\sigma_{ess}|A|$.
\end{lem}

\noindent{\bf Proof.} Let $C=|A+B|^2$, defined on $|A|^{-2}\h$. We
show that $(C-\zeta^2)^{-1}$ is a meromorphic operator valued
function on $\zeta\in \mathbb{C} \backslash [\sigma(|A|) \cup
\sigma(-|A|)]$. This follows from the identity
\begin{equation}\label{identity}
(C-\zeta^2)^{-1}=(|A|+\zeta)^{-1}[1+N(\zeta)]^{-1}(|A|-\zeta)^{-1}
\end{equation}
where
$$N(\zeta)=(|A|-\zeta)^{-1}[A^\ast B+B^\ast A+B^\ast
B](|A|+\zeta)^{-1}$$ is an analytic family of compact operators on
$\zeta\in \mathbf{C} \backslash [\sigma(|A|) \cup
\sigma(-|A|)]$.$\square$

\medskip

\noindent{\bf Proof of Theorem 4.1.}{\it i)} Taking
$A=-e^{-2\theta}{\bf \nabla}^2-z$ and $B=v_\theta+\gamma w_\theta$,
it follows that the essential spectrum of $|H_\theta(\gamma)-z|$ is
contained in $\sigma(|-e^{-2\theta}{\bf \nabla}^2-z|)$, which is
$[d(z,\theta),\infty)$.

{\it ii)} We need to show that $|H_\theta(\gamma)-z|$ has spectrum
below $d(z,\theta)$. Let $\psi_0$ be the eigenvector corresponding
to the resonance, $H_\theta\psi_0=z_0 \psi_0$. Remark that
$$(H_\theta(\gamma)-z)\psi_0=(z_0-z)[1+\gamma w_\theta
(H_\theta-z)^{-1}]\psi_0$$ and consequently
$$\||H_\theta(\gamma)-z|\psi_0\|\leq|z_0-z|(1+\gamma \| w_\theta
(H_\theta-z)^{-1}\|\, ).$$ With our assumptions, there exists $0<a<1$
and $b>0$ such that $\|w_\theta \psi\|\leq a \|{\bf \nabla}^2
\psi\|+b\|\psi\|$ and similarly for $v_\theta$, for any $\psi\in
\mathcal{D}({\bf \nabla}^2)$.  A relatively elementary
manipulation then yields:
$$\|w_\theta (H_\theta-z)^{-1}\|\leq
\frac{a}{1-a}+\frac{b+a|z|}{1-a}\|(H_\theta-z)^{-1}\|.$$ The
conclusion is that we can make $\||H_\theta(\gamma)-z|\psi_0\|$
arbitrarily small, in particular, smaller than $d(z,\theta)$, by
taking the limit $z\rightarrow z_0$ and $\gamma\rightarrow 0$.
Consequently, $\inf \sigma (|H_\theta(\gamma)-z|)<d(z,\theta)$ for
$\gamma$ small enough and $z$ close enough to the resonance.

{\it iii)} Since the operator $H_\theta(\gamma)-z$ is
$C$-selfadjoint, it admits the decomposition stated by Theorem 2.4:
$$ H_\theta(\gamma)-z = CJ|H_\theta(\gamma)-z|,$$
where the second conjugation $J$ commutes, in the strong sense,
with the selfadjoint operator $|H_\theta(\gamma)-z|$. In
particular $J$ leaves invariant the spectral subspaces of
$|H_\theta(\gamma)-z|$. Thus if $\lambda_n$ belongs to the
discrete spectrum of $|H_\theta(\gamma)-z|$, then the vector space
consisting of the eigenvectors $\phi_n \in {\mathcal D}({\bf
\nabla}^2)$:
$$ |H_\theta(\gamma)-z| \phi_n = \lambda_n \phi_n$$
is left invariant by $J$. Thus, either $\phi_n = -J\phi_n$ or
$\phi'_n = \phi_n + J \phi_n$ provide a new eigenvector $\psi_n$
satisfying $J \psi_n = \psi_n$. Therefore,
$$ C (H_\theta(\gamma)-z) \psi_n = J |H_\theta(\gamma)-z| \psi_n =|H_\theta(\gamma)-z| \psi_n=  \lambda_n \psi_n.$$ This proves the
last assertion in the statement.$\square$

\section{Conclusions}

The goal of this paper was to introduce recent results in theory of
complex symmetric operators and to show, through two non-trivial
examples, their potential usefulness in the study of Schr\"odinger
operators. It was shown that, for complex selfadjoint operators with
compact resolvent, a certain antilinear eigenvalue problem can be
used in a similar way the spectral decomposition for self-adjoint
operators is used to estimate the norm of the resolvent. We combined
this observation with the complex scaling technique, to obtain sharp
lower bounds on the exponential decay of the resolvent and density
matrix for Schr\"odinger operators with a gap in the energy spectrum.
We also shown that these techniques can be applied to complex
selfadjoint operators with non-compact resolvent, in particular, to
Schr\"odinger appearing in the complex scaling theory of resonances.

In concrete applications, we believe that one can go beyond the approximations 
made in this paper and use either the newly developed min-max principle for complex-symmetric operators \cite{Danciger,DGP}, or direct, numerical integration of the anti-linear eigenvalue problem to obtain sharper, problem specific estimates of the exponential decay constant or to evaluate the resolvent norm in our second application. 

\medskip

{\bf Acknowledgment.} Paper supported by the National Science
Foundation Grants No. DMS 0100367 and DMR 0313980.

\end{document}